\documentclass{iauc}
\usepackage{graphicx}

\title[IntraCluster Light at $z\sim0.25$] %% give here short title %% 
{Diffuse light in $z\sim0.25$ galaxy clusters:
constraining tidal damage and the faint end of the Luminosity Function}

\author[S.~Zibetti \& S.D.M.~White]   %% give here short author list %%
{Stefano Zibetti$^{1}$ %
\and Simon D.M. White$^2$}

\affiliation{$^1$Max-Planck-Institut f\"ur Extraterrestrische Physik \break
Gie\ss enbachstra\ss e, D-85748, Garching bei M\"unchen, Germany \break 
email: szibetti@mpe.mpg.de\\[\affilskip]
$^2$Max-Planck-Institut f\"ur Astrophysik %\break
$-$ Garching bei M\"unchen, Germany %\break 
}

\pubyear{2005}
\volume{198} %% insert here IAU Symposium No.
\pagerange{0--3}
\date{?? and in revised form ??}
\jname{Proceedings Title IAU Colloquium }
\editors{B. Binggeli, and H. Jerjen, eds.}
\begin{document}

\maketitle
\begin{abstract}
The starlight coming from the intergalactic space in galaxy clusters and groups
witnesses the violent tidal interactions that galaxies experience in these dense
environments. Such interactions may be (at least partly) responsible for the
transformation of normal star-forming galaxies into passive dwarf ellipticals
(dEs).\\ In this contribution we present the first systematic study of the
IntraCluster Light (ICL) for a statistically representative sample
(Zibetti et al. 2005), which comprises 683  clusters selected between  z=0.2
and 0.3 from $\sim1500 \deg^2$ in the SDSS. Their ICL is studied by stacking the
images in the $g$-, $r$-, and $i$-band after masking out all galaxies and
polluting sources. In this way a very uniform background illumination is
obtained, that allows us to measure surface brightnesses as faint as 31 mag
arcsec$^{-2}$ and to trace the ICL out to 700 kpc from the central galaxy. We
find that the local fraction of light contributed by intracluster stars rapidly
decreases as a function of the clustercentric distance, from $\sim$40\% at 100
kpc to $\sim$5\% at 500 kpc. By comparing the distribution and colours of the
ICL and of the clusters galaxies, we find indication that the main source of ICL
are the stars stripped from galaxies that plunge deeply into the cluster
potential well along radial orbits. Thus, if dEs are the remnants of these
stripped progenitors we should expect similar radial orbital anisotropies and
correlations between the dE luminosity  function and the amount of ICL in
different clusters.\\
The diffuse emission we measure is contaminated by faint unresolved galaxies:
this makes our flux estimate depend to some extent on the assumed luminosity
function, but, on the other hand, allows us to constrain the number of faint
galaxies. Our present results disfavour steep ($\alpha<-1.35$) faint-end 
powerlaw slopes.

%% add here a maximum of 10 keywords, to be taken form the file <Keywords.txt>.

\end{abstract}

\firstsection % if your document starts with a section,
\section{Introduction} 
The diffuse intracluster light (ICL) is produced by stars which are not bound
to any individual galaxy in a cluster, but orbit freely in the global
gravitational potential of the cluster. The ICL has now been detected in a
number of nearby groups and clusters both by means of deep surface photometry
\cite[(e.g. Bernstein et al. 1995; Feldmeier et al. 2002; Gonzalez, Zabludoff
\& Zaritsky 2005)]{bernstein+95,feldmeier+02,gonzalez+05} and by  resolving red
giant stars \cite[(Durrell et al. 2002)]{durrell+02} and planetary nebulae 
\cite[(e.g. Arnaboldi et al. 1996)]{arnaboldi+96}. As for the origin of this
stellar component, the presence of tidal features and  dynamical substructure,
that has emerged from a number of studies, strongly supports  the hypothesis
that intergalactic stars are created via tidal stripping from galaxies and
during merger events, rather than \emph{in situ} from isolated extragalactic
star forming regions.\\  Tidal stripping and repeated tidal perturbations in a
dense environment (harassment) may be also invoked as the main mechanism
responsible for transforming low mass star-forming discs into dwarf elliptical
galaxies (dEs) and compact dwarves. The investigation of the systematic
properties of the ICL puts quantitative constraints on the tidal damage that
galaxies suffer in clusters and, therefore, can provide invaluable clues to
understand the formation of dEs.\\
\cite[Zibetti et al. (2005)]{zibetti_etal05} have recently analysed the
systematic properties of the diffuse  stellar emission in a sample of 683
clusters by stacking their images from the Sloan Digital Sky Survey 
\cite[(SDSS, York et al. 2000)]{SDSS}. In this contribution we will
review their results, focusing on the implications for the stripping scenario
for the formation of dEs.

\section{Sample and photometric technique}
The stacking technique aims at obtaining high photometric sensitivity, while
strongly reducing the systematic defects that arise in individual observations
from flat field inhomogeneities and scattered light from ``polluting'' (fore-
and background) sources. This is achieved by averaging a large number of images
of different targets, where the unwanted sources are efficiently masked. Typical
required number of images is several hundreds, hence the need for a large
imaging database as the SDSS. The results obtained from the stacking of a large
number of clusters grant a high statistical significance, but have the
disadvantage of lacking information about  substructure and cluster-to-cluster
variations, that must be derived from the analysis of different subsamples.

For the present study a sample of 683 clusters of galaxies has been drawn from 
$\sim$1,500 deg$^2$ of the first data release of the SDSS 
\cite[(Abazajian et al. 2003)]{DR1}. We have used the
max-BCG cluster finder algorithm developed by J. Annis to select clusters in the
same richness range as the Abell catalogue, between $z=0.2$ and 0.3. The redshift
constraint has been introduced to limit the size of the stacked images to a
single SDSS frame, thus avoiding to mosaic images with different background 
properties for a given cluster. Moreover, at the median $z$ of the sample the
$g-r$ colour conveniently maps the 4000\AA~break, providing valuable information
about the intracluster stellar population.

Lacking any kinematic characterisation, we \emph{define} as ICL all the flux
beyond the isophote 25 mag arcsec$^{-2}$ (observed $r$-band) of any galaxy. All
galaxies, except the central brightest cluster galaxy (BCG), are masked to this
extent. Bright stars are masked out to 3 times the isophotal radius, to prevent
their bright haloes being included in the stacking. Images are then centred on
the BCG, size-rescaled to the same metric radius, and intensity-rescaled to the
same photometric calibration, corrected for Galactic extinction and cosmological
SB dimming. A simple average of the masked images is then computed (stacked
image). A stacking of images where only the brightest stars were masked has been
performed as well, to obtain a map of the total luminosity of the cluster to be
compared to the ICL.

We have tested the efficiency of our masking algorithm and quantified the
fraction of galaxy light which fails to be masked. Realistic SB distribution
have been used to simulate galaxies in the same observed conditions as for real
clusters, with different luminosity functions (LFs) parametrised \'a la
Schechter. Depending on the adopted parameters (and particularly on the
powerlaw slope of the faint end $\alpha$) we find that 5 to 20-25\% of the light
in galaxies can be missed. As we will show below, this effect can be used along
with other observational constraints to put limits to the faint end slope of
the LF.

The following \emph{results} are derived assuming that the fraction of
unmasked galaxy light is 15\%, as obtained for the 
parameters of the Coma cluster LF ($M_r^*=-21.37$, $\alpha=-1.18$,
\cite[Mobasher et al. 1998]{comaLF}), passively evolved to $z=0.25$.

\section{Results} 
\begin{figure}
%\begin{centerline}
\includegraphics[width=0.5\textwidth]{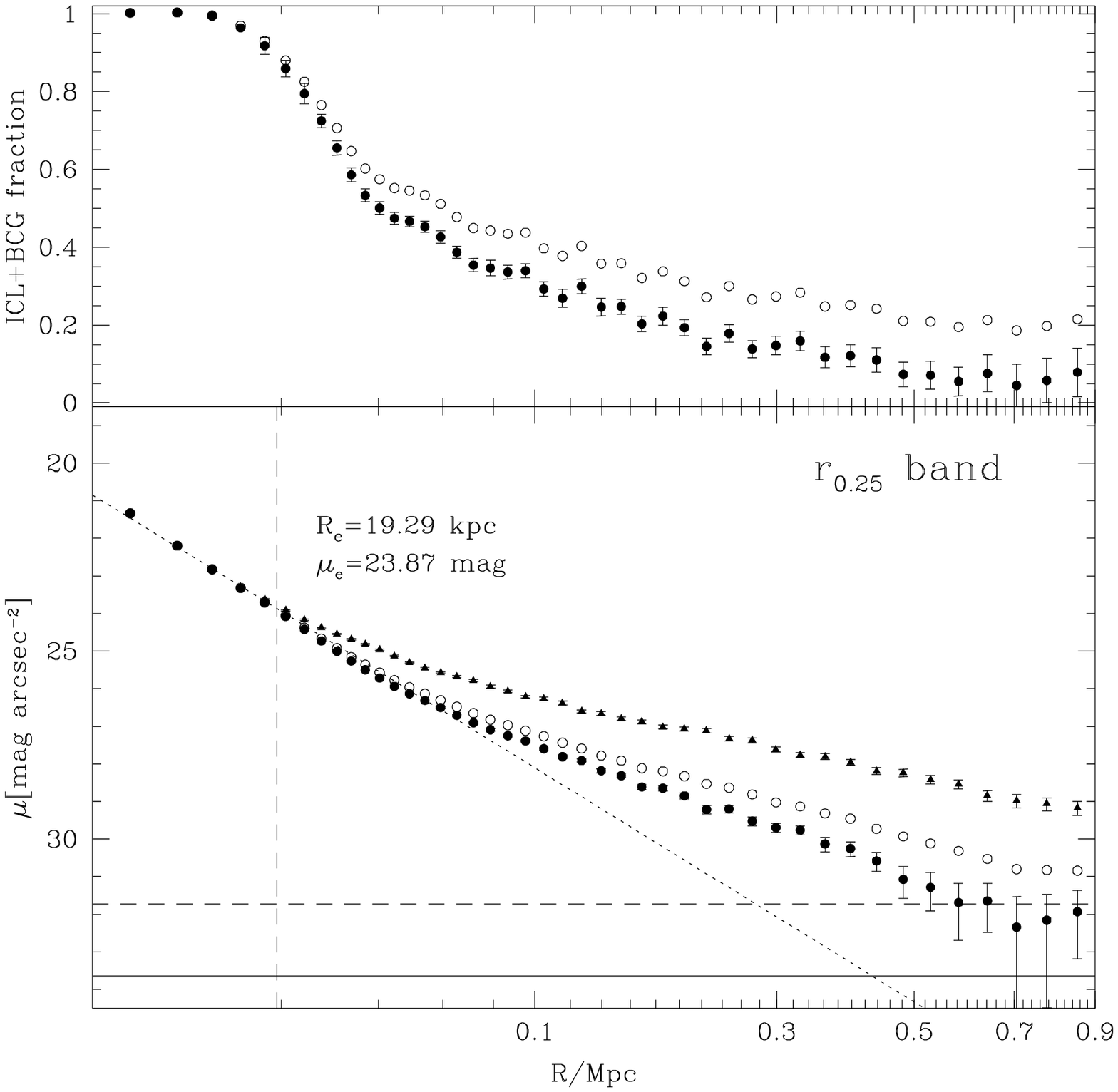}
\includegraphics[width=0.5\textwidth]{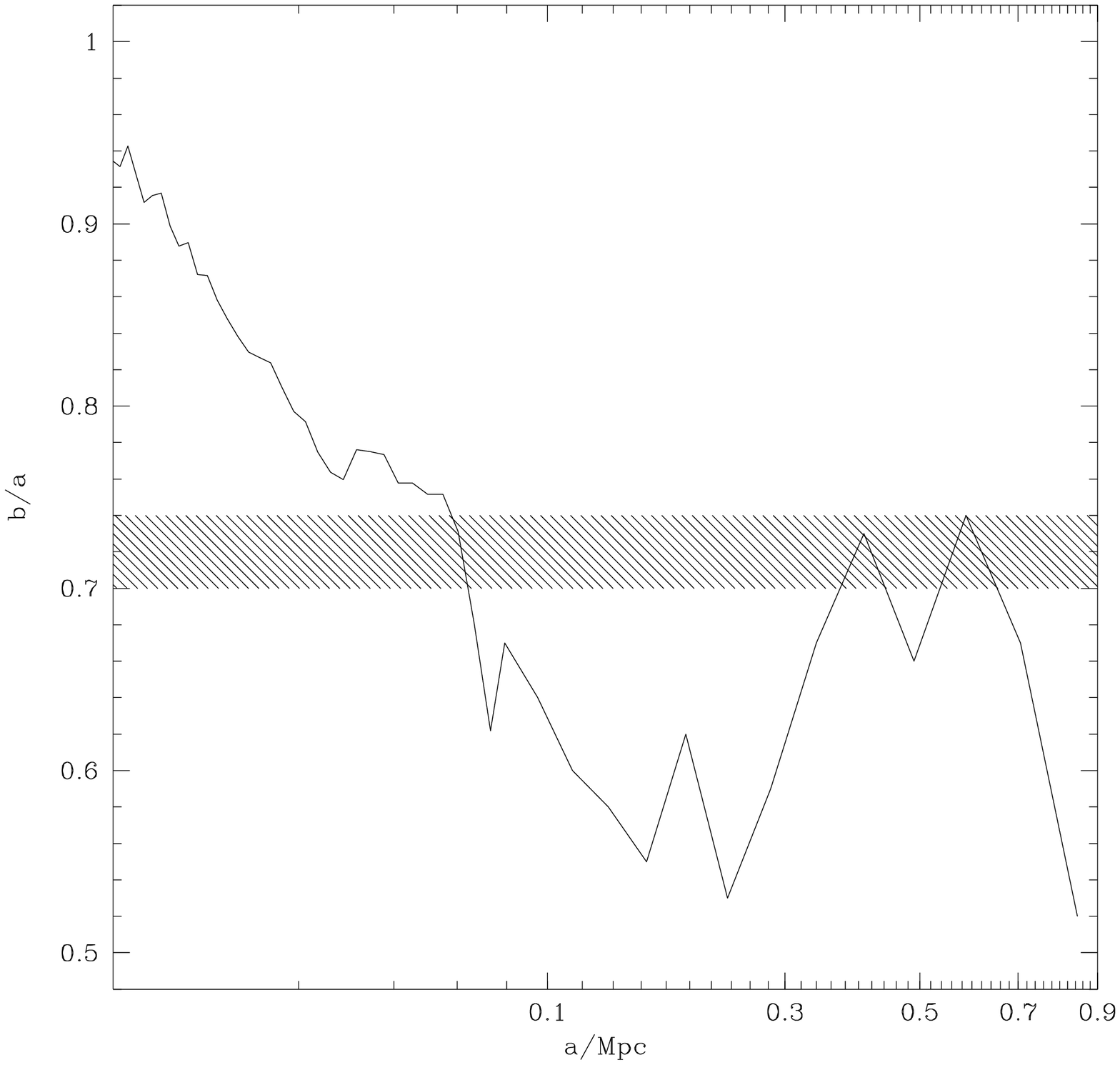}
%\end{centerline}
\caption{Panel a): $r$-band surface photometry. The lower section shows the
surface brightness profiles for all components (triangles with error bars), for
the diffuse light as measured (open circles) and for the corrected ICL (filled
circles with error bars). The dotted line represents the best fitting de
Vaucouleurs profile to the BCG core, with the vertical dashed line marking
$R_e$. The two horizontal lines show 1$\sigma$ background uncertainty for the
total (dashed) and for the diffuse light. The upper section displays the local
relative amount of diffuse light (open circles) and of the corrected ICL+BCG
component (filled circles with error bars). Panel b): Isophotal shapes for
clusters with a flattened BCG ($b/a<0.7$ from 2-D de Vaucouleurs fit). The solid
line displays the axial ratio of the (uncorrected) diffuse light + BCG, with
typical errors of 0.05 in the central regions to 0.1 at 500 kpc. The shaded
area is the 1$\sigma$ interval for the distribution of the total
light.}\label{profiles}
\end{figure}

By extracting azimuthally averaged SB profiles we can trace the diffuse light
out to $\sim$700 kpc from the central galaxy at levels of 31 mag arcsec$^{-2}$.
In the lower panel of fig. \ref{profiles}a) the combined contribution of the
BCG and of the ICL is shown by filled circles with error bars. Starting from 80
kpc, the ICL clearly appears as an excess with respect to the inner de
Vaucouleurs profile that characterises the BCG. Compared to the total light of
all stellar components (triangles), the ICL is remarkably more centrally
concentrated, providing a local relative contribution that decreases from 40\%
of the total SB at 100 kpc, to less than 5\% at 600 kpc and beyond. This
clearly indicates that the creation/accumulation of intracluster stars is much
more efficient in the centre of the cluster. The colour profiles of the ICL
($g-r$ and $r-i$, obtained from the $g$-, $r$- and $i$-band stacked images) are
consistent with ($g-r$) or just marginally redder than ($r-i$) the average
colours of galaxies, thus suggesting that the intracluster stellar population
is not dissimilar from that in galaxies.
On a subsample of clusters with BCG having a significant flattening ($b/a<0.7$
from 2D de Vaucouleurs SB fitting), the stacking has been performed after
aligning the images along the major axis of the BCG. The isophotal analysis
(fig. \ref{profiles}b) shows that the ICL is strongly aligned with the BCG and
displays even a higher degree of flattening. In fact, there is a monotonic
decrease of $b/a$ from the BCG core out to 300 kpc, where $b/a\sim 0.5$. It is
remarkable that the distribution of galaxies is less aligned with the BCG
and/or less flattened than the ICL. This may indicate that the intracluster stars
are stripped more efficiently from galaxies that move along radial orbits
aligned with the main axis of the BCG.

We have studied the dependence of the ICL SB and luminosity on cluster richness
and BCG luminosity. The SB of the ICL is enhanced in rich clusters and in those
having a luminous BCG, and \emph{vice versa}. The clusters with the faintest
BCGs appear particularly deficient in ICL. However, the relative contribution
of ICL to the total luminosity within 500 kpc is roughly constant 11\% for all
clusters.

\section{dEs as stripped galaxies?}
N-body hydrodynamical simulations (also Mastropietro in this Colloquium)
have demonstrated that dynamical harassment, tidal and ram-pressure stripping
can transform small star-forming disc galaxies into passive dEs
\cite[(see also Merritt 1984, Moore et al. 1996)]{merritt84,moore+96}. It is
therefore conceivable that dEs and the ICL share the same origin. The broadband
colours are consistent with this scenario, and the fact that the total
luminosity of dEs is similar to that of the ICL is very intriguing too. On the
other hand, if dEs are the remnants of the galaxies from which a large number of
stars have been stripped into the intergalactic space, we should expect (i) that
in anisotropic clusters dEs display a similar flat and aligned distribution as
the ICL, in  addition to some degree of radial orbital anisotropy; and (ii)
that the luminosity function of dEs correlates with the amount of ICL. The
study of these correlations between ICL and dEs may therefore provide
invaluable clues on the origins of these two components and on the violent 
history of galaxy clusters.

\section{Constraints to the faint end of the LF}
In all the analysis conducted above, the measured diffuse light is converted
into ``real'' ICL assuming that the fraction of galaxy light that is left
unmasked is 15\%, according to our simulations for a LF as in Coma. As already
mentioned, by varying the parameters of the LF values between 5 and 20-25\% are
found. However we have a number of constraints that limit the range of
variation to $<$5\% around the assumed 15\%. $M^*$ is constrained within a few
hundredths of mag by directly observed number counts. On the other hand, if a
too steep LF is chosen, then the predicted unmasked flux becomes larger than the
actual flux in the diffuse component. In particular, if we assume that the LF is
independent of radius, then the minimum value of the local fraction of diffuse 
light represents the maximum fraction of light that might be missed by our masks.
Since this fraction is 20\% at $R\sim600$ kpc (see fig. \ref{profiles}a), we 
calculate that the faint-end powerlaw of a simple Schechter LF cannot be steeper
than $\alpha=-1.35$.


\begin{thebibliography}{}
\bibitem[\protect\citeauthoryear{Abazajian et 
al.}{2003}]{DR1} Abazajian K., et al., 2003, \textit{AJ}, 126, 2081
\bibitem[\protect\citeauthoryear{{Arnaboldi}, {Freeman}, {Mendez},
  {Capaccioli}, {Ciardullo}, {Ford}, {Gerhard}, {Hui}, {Jacoby}, {Kudritzki} \&
  {Quinn}}{{Arnaboldi} et~al.}{1996}]{arnaboldi+96}
{Arnaboldi} M.,  {Freeman} K.~C.,  {Mendez} R.~H.,  {Capaccioli} M.,
  {Ciardullo} R.,  {Ford} H.,  {Gerhard} O.,  {Hui} X.,  {Jacoby} G.~H.,
  {Kudritzki} R.~P.,    {Quinn} P.~J.,  1996, \textit{ApJ}, 472, 145
\bibitem[\protect\citeauthoryear{{Bernstein}, {Nichol}, {Tyson}, {Ulmer} \&
  {Wittman}}{{Bernstein} et~al.}{1995}]{bernstein+95}
{Bernstein} G.~M.,  {Nichol} R.~C.,  {Tyson} J.~A.,  {Ulmer} M.~P.,
  {Wittman} D.,  1995, \textit{AJ}, 110, 1507
\bibitem[\protect\citeauthoryear{{Durrell}, {Ciardullo}, {Feldmeier}, {Jacoby}
  \& {Sigurdsson}}{{Durrell} et~al.}{2002}]{durrell+02}
{Durrell} P.~R.,  {Ciardullo} R.,  {Feldmeier} J.~J.,  {Jacoby} G.~H.,
  {Sigurdsson} S.,  2002, \textit{ApJ}, 570, 119
\bibitem[\protect\citeauthoryear{{Feldmeier}, {Mihos}, {Morrison}, {Rodney} \&
  {Harding}}{{Feldmeier} et~al.}{2002}]{feldmeier+02}
{Feldmeier} J.~J.,  {Mihos} J.~C.,  {Morrison} H.~L.,  {Rodney} S.~A.,
  {Harding} P.,  2002, \textit{ApJ}, 575, 779
\bibitem[\protect\citeauthoryear{Gonzalez, Zabludoff, \& 
Zaritsky}{2005}]{gonzalez+05} Gonzalez A.~H., Zabludoff A.~I., 
Zaritsky D., 2005, \textit{ApJ}, 618, 195  
\bibitem[\protect\citeauthoryear{{Merritt}}{{Merritt}}{1984}]{merritt84}
{Merritt} D.,  1984, \textit{ApJ}, 276, 26
\bibitem[\protect\citeauthoryear{{Moore}, {Katz}, {Lake}, {Dressler} \&
  {Oemler}}{{Moore} et~al.}{1996}]{moore+96}
{Moore} B.,  {Katz} N.,  {Lake} G.,  {Dressler} A.,    {Oemler} A.,  1996,
  \textit{Nature}, 379, 613
\bibitem[\protect\citeauthoryear{York et al.}{2000}]{SDSS} 
York D.~G., et al., 2000, \textit{AJ}, 120, 1579 
\bibitem[\protect\citeauthoryear{Zibetti et 
al.}{2005}]{zibetti_etal05} Zibetti S., White S.~D.~M., Schneider 
D.~P., Brinkmann J., 2005, \textit{MNRAS}, 358, 949 

\end{thebibliography}
\end{document}